\documentclass[9pt,twocolumn]{article}%
\usepackage{url}
\usepackage{amsmath}
\usepackage{subfig}
\usepackage{amssymb}
\usepackage[conf]{ametsoc}
\usepackage{graphicx}
\usepackage{amsfonts}%
\renewcommand{\cite}[2][]{\citep[#1]{#2}}
\begin{document}

\title{Ignition from a Fire Perimeter in a WRF Wildland Fire Model\footnotemark[1]}
\author{Volodymyr Y. Kondratenko\footnotemark[2], Jonathan D. Beezley\footnotemark[2],
Adam K. Kochanski\footnotemark[3], and Jan Mandel\footnotemark[2]}
\maketitle

\begin{abstract}
The current WRF-Fire model starts the fire from a given ignition point at a
given time. We want to start the model from a given fire perimeter at a given
time instead. However, the fuel balance and the state of the atmosphere depend
on the history of the fire. The purpose of this work is to create an
approximate artificial history of the fire based on the given fire perimeter
and time and an approximate ignition point and time. Replaying the fire
history then establishes a reasonable fuel balance and outputs heat fluxes
into the atmospheric model, which allow the atmospheric circulation to
develop. Then the coupled atmosphere-fire model takes over. In this
preliminary investigation, the ignition times in the fire area are calculated
based on the distance from the ignition point to the perimeter, assuming that
the perimeter is convex or star-shaped. Simulation results for an ideal
example show that the fire can continue in a natural way from the perimeter.
Possible extensions include algorithms for more general perimeters and running
the fire model backwards in time from the perimeter to create a more realistic
history. The model used extends WRF-Fire and it is available from \url{openwfm.org}.

\end{abstract}

\footnotetext[1]{WRF Summer Workshop 2011. This research was supported by the
National Science Foundation under grant AGS-0835579, and by U.S. National
Institute of Standards and Technology Fire Research Grants Program grant
60NANB7D6144.} \footnotetext[2]{Department of Mathematical and Statistical
Sciences, University of Colorado Denver, Denver, CO} \footnotetext[3]%
{Department of Meteorology, University of Utah, Salt Lake City, UT}

\section{Introduction}

Fire models generally start the fire from a given ignition point at a given
time, and sophisticated ignition parameterizations exist, including line
ignition and submesh ignition procedures \cite{Mandel-2011-CAF}. However, in
practice one is often faced with the need to start a fire model from a fire
already in progress. Often only perimeter data pertaining to some time are
available, such as from the US\ Forest Service at
\url{activefiremaps.fs.fed.us}. This need arizes in analyses of existing
fires currently, and it will become even more important for forecasting of the
behavior of fires in progress in future.

With models that do not include two-way interaction with the atmosphere,
continuing from an existing fire state is essentially straightforward. While
the fuel can be partially burned in some areas,\ fuel in locations untouched
by the fire is unchanged, and the model can simply progress to the new areas
regardless of the fire history. (A model that may include long-range effects
such as preheating in front of the fire, would be an exception.) In a coupled
atmosphere-fire model, however, the situation is very different. First, simply
igniting the whole area inside the given perimeter is not an option, because
the large instantaneous heat release will cause the model to break down. More
importantly, the state of the atmosphere evolves in interaction with the fire,
and the buyoancy caused by the heat flux causes significant changes to the
wind field, which in turn influences the future progress of the fire.

Starting the fire from a state already developed is essentially a data
assimilation problem, and it could be treated as such by a shooting
method:\ the fire starts from a ignition point at a time in the past, then at
the given simulation time, the state of the fire is compared with the given
perimeter, and adjustments can be made to the ignition time and location, much
as in variational data assimilation methods such as 4DVAR. We plan to study
such approaches in future as a part of our effort in the area of data
assimilation for wildland fires
\cite{Beezley-2009-HDA,Beezley-2008-MEK,Mandel-2008-WFM,Mandel-2009-DAW,Mandel-2010-FFT}%
.

The approach adopted here is different. Given a fire perimeter, we create an
artificial history. Then the fire history is replayed, which produces heat
output into the atmosphere, and the atmospheric model spins up to a state that
is plausible for the fire at the stage given by the perimeter. The artificial
history is essentially a parameterization of the process that leads to the
development of the fire perimeter. Parameterizations of various levels of
sophistication can be considered, up to and including running a fire model
backwards in time to find an ignition point and iterating to find a matching
atmospheric state. In this initial study, we consider a very simple artificial
history model, and show that it results in acceptable fire and atmosphere
states for the perimeter. Approximate perimeter states obtained by such method
could provide also a good starting point for data assimilation in future.

\section{The model}

Fire models range from tools based on \citet{Rothermel-1972-MMP} fire spread
rate formulas, such as BehavePlus \citep{Andrews-2007-BFM} and FARSITE
\citep{Finney-1998-FFA}, suitable for operational forecasting, to
sophisticated \mbox{3-D} computational fluid dynamics and combustion
simulations suitable for research and reanalysis, such as FIRETEC
\citep{Linn-2002-SWB} and WFDS \citep{Mell-2007-PAM}. BehavePlus, the PC-based
successor of the calculator-based BEHAVE, determines the fire spread rate at a
single point from fuel and environmental data; FARSITE uses the fire spread
rate to provide a \mbox{2-D} simulation on a PC; while FIRETEC and WFDS model
combustion in 3D, which is much more expensive. See the survey by
\citet{Sullivan-2009-RWF} for a number of other models.

The model considered here couples the mesoscale atmospheric code
WRF-ARW~\cite{Skamarock-2008-DAR} with a fire spread module, based on the
Rothermel model \cite{Rothermel-1972-MMP} and implemented by the level set
method. In each time step, the fire model inputs the atmospheric winds and
outputs surface sensible and latent heat fluxes into the atmosphere. Only the
finest domain in WRF\ is coupled with the fire model. The fire model works in
conjunction with WRF land use models, and it interpolates horizontal winds from the
ideal logarighmic wind profile to appropriate heights above the surface, for each fuel.

The model has grown out of NCAR's CAWFE code
\cite{Clark-1996-CAM,Clark-1996-CAF-a,Clark-2004-DCA,Coen-2005-SBE}, which
couples the Clark-Hall atmospheric model with fire spread implemented by
tracers, and it got its start from a prototype code coupling the fire model in
CAWFE with WRF in LES mode \cite{Patton-2004-WCA}. The tracers, however, were
replaced by a level set method, which we considered more flexible and more
suitable for data assimilation and WRF parallel infrastructure. The coupled
model is capable of running faster than real time in LES mode, with resolution
of tens of meters for the atmosphere, and meters for the fire, with the
matching time step of a fraction of a second, on the innermost modeling domain
of many kilometers in size \cite{Jordanov-2011-SWF}. Fuel data and topography
can be obtained from government databases in the United States and from
satellite images and GIS elsewhere \cite{Jordanov-2011-SWF}. See
\citet{Mandel-2009-DAW,Mandel-2011-CAF} for futher details and references. The
model is currently available from the Open Wildland Fire Modeling environment
at \url{openwfm.org}, along with utilities for data preparation,
visualization, and diagnostics and a wiki with many user guides for the
specific features and utilities, discussion, and support. A code containing a
subset of the features is distributed with WRF as WRF-Fire.

\section{Encoding and replaying the fire history}

The state of the fire model consists of a level set function, $\Phi$, given by
its values on the nodes of the fire model mesh, and time of ignition
$T_{\mathrm{i}}$. The level set function is interpolated linearly. At a given
simulation time $t$, the fire area is the set of all points $\left(
x,y\right)  $ where $\Phi\left(  t,x,y\right)  \leq0$. The level set function
and the ignition time satisfy the consistency condition%
\begin{equation}
\Phi\left(  t,x,y\right)  \leq0\Longleftrightarrow T_{\mathrm{i}}\left(
x,y\right)  \leq t, \label{eq:consistency}%
\end{equation}
as both of these inequalities express the condition that the location $\left(
x,y\right)  $ is burning at the time $t$. In every time step of the
simulation, the level set function is advanced by one step of a Runge-Kutta
scheme for the level set equation%
\[
\frac{d\Phi}{dt}=-R\left\Vert \nabla\Phi\right\Vert ,
\]
where $R=R\left(  t,x,y\right)  $ is the fire rate of spread, which depends on
the fuel, wind speed, and slope. The ignition time at nodes is then computed
for all newly ignited nodes, and it satisfies the consistency condition
(\ref{eq:consistency}).

The fire history is encoded as an array of ignition times $T_{\mathrm{i}%
}\left(  x,y\right)  $, prescribed at all fire mesh nodes. To replay the fire in
the period $0\leq t\leq T_{\mathrm{per}}$, the numerical scheme for
advancing $\Phi$ and $T_{\mathrm{i}}$ is suspended, and instead the level set
function is set to%
\[
\Phi\left(  t,x,y\right)  =T_{\mathrm{i}}\left(  x,y\right)  -t.
\]
After the end of the replay is reached, the numerical scheme of the level set
method is started from the level set function $\Phi$ at
$t=T_{\mathrm{per}}$.

For reasons of numerical accuracy and stability, the level set function needs
to have approximately uniform slope. For example, a very good level set
function, which has slope equal to one, is the signed distance from a given
closed curve $\Gamma$,
\[
\Phi\left(  x,y\right)  =\pm\operatorname*{dist}\left(  \left(  x,y\right)
,\Gamma\right)  ,
\]
where the sign is taken to be negative inside the region limited by $\Gamma$,
and positive outside \cite{Osher-2003-LSM}. Thus, the ignition times
$T_{\mathrm{i}}$ need to be given on the whole domain and they need to be such
that $T_{\mathrm{i}}$ decreases with the distance from the given perimeter
inside the fire region, and increases outside. The ignition times
$T_{\mathrm{i}}$ outside of the given fire perimeter are perhaps best thought
of as what the ignition times might be in future as the fire keeps burning.

\section{Creating an artificial fire history}

The purpose of this algorithm is to create the artificial values of the time
of ignition on the fire model mesh, given ignition point ($x_{\mathrm{ign}}$,
$y_{\mathrm{ign}}$), ignition time $T_{\mathrm{ign}}$, fire perimeter $\Gamma
$, and the time when the fire reached the perimeter $T_{\mathrm{per}}$,
assuming that the fire perimeter is convex, or at least star-shaped with respect
to the ignition point. The fire perimeter
is given as a set of points $(x_{k},y_{k})$ in the fire model domain,
$k=1,\ldots,n$  which form a closed curve consisting of line segments $\left[
(x_{k},y_{k}),(x_{k+1},y_{k+1})\right]  $ between each two successive points.
We take $(x_{1},y_{1})=(x_{n+1},y_{n+1})$ so that the starting and the ending
point are identical. The coordinates of the point of ignition and of the
points defining of the fire perimeter do not need to coincide with mesh
points of the grid.

The method consists of linear interpolation of the ignition time between
$T_{\mathrm{ign}}$ at the ignition point and $T_{\mathrm{per}}$ on the
perimeter, along straight lines connecting the ignition point with points on
the perimeter. The ignition time is also extrapolated beyond the perimeter in
the same manner to provide a suitable level set function, as discussed in the
previous section. Given a mesh point with coordinates $\left(  x,y\right)  $,
the algorithm to determine the ignition time $T_{\mathrm{i}}\left(
x,y\right)  $ consists of the following steps.

\begin{figure}[t]
\begin{center}
\includegraphics[height=2.5in]{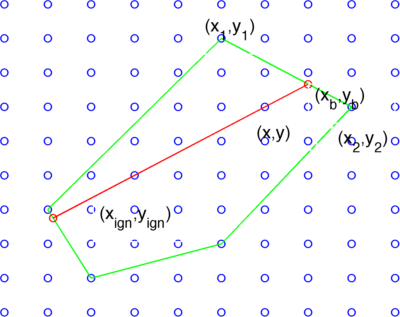}
\end{center}
\caption{Construction of intersection of the fire perimeter and the half line originating from the
Ignition point and passing through a given mesh point.}%
\label{fig:perimeter}%
\end{figure}

\begin{enumerate}
\item Find the intersection $\left(  x_{\mathrm{b}},y_{\mathrm{b}}\right)  $
of the fire perimeter and the half-line starting at the ignition point and
passing through the point $\left(  x,y\right)  $ 
(Fig.~\ref{fig:perimeter}).
For this
purpose, we use the function%
\begin{equation*}
\hspace{-.15in}F(x,y,x_{b},y_{b})=(y_{b}-y_{\mathrm{ign}})(x-x_{\mathrm{ign}})-(x_{b}%
-x_{\mathrm{ign}})(y-y_{\mathrm{ign}}),
\end{equation*}
which is zero if point $\left(  x_{\mathrm{b}},y_{\mathrm{b}}\right)  $ lies
on the line defined by $(x,y)$ and $(x_{\mathrm{ign}},y_{\mathrm{ign}})$, and
it is positive in one half-plane and negative in the other. We then find
segment $\left[  (x_{k},y_{k}),(x_{k+1},y_{k+1})\right]  $ such that
$F(x,y,x_{k},y_{k})F(x,y,x_{k+1},y_{k+1})<0$, that is, the points $(x_{k}%
,y_{k})$ and $(x_{k+1},y_{k+1})$ lie on opposite sides of the line passing
through $(x,y)$ and $(x_{\mathrm{ign}},y_{\mathrm{ign}})$. Since the line
intersects the fire perimeter at two points, one on each side of the ignition
point, we choose correct segment as follows:

\begin{itemize}
\item If $\left(  x,y\right)  $ is inside $\Gamma$, that is, closer to the
ignition point than to the intersection, then the desired segment is the one
that lies on the same side from the ignition point as the point $\left(
x,y\right)  $; 

\item If $\left(  x,y\right)  $ is outside of $\Gamma$, then the needed
segment lies on the same side from the ignition point as $\left(  x,y\right)  $.
\end{itemize}

\item Calculate the time of ignition of the mesh point, based on the ratio of
the distances of the mesh point and the perimeter point to the 
ignition point,

\[
T_{\mathrm{i}}(x,y)=T_{\mathrm{ign}}+\frac{\left\Vert \left(  x,y\right)
-\left(  x_{\mathrm{ign}},y_{\mathrm{ign}}\right)  \right\Vert }{\left\Vert
\left(  x_{\mathrm{b}},y_{\mathrm{b}}\right)  -\left(  x_{\mathrm{ign}%
},y_{\mathrm{ign}}\right)  \right\Vert }\left(  T_{\mathrm{per}%
}-T_{\mathrm{ign}}\right).
\]

\end{enumerate}

\section{Computational results}

We have  tested this algorithm on an ideal example to measure the difference in the atmospheric winds between a simulation
propagated naturally from a point and another one advanced artificially.
In this example, the topography was flat except for a small hill roughly $500$\,m in diameter and $100$\,m high in the center of a domain
of size $2.4$\,km $\times$ $2.4$\,km.  The atmospheric and fire grid resolutions used were $60$ m and $6$ m respectively, with a 
$0.25$\,s time step.  The background winds were approximately $9.5$\,m/s traveling southwest at the
lowest atmospheric layer $30$\,m above the surface.  The first simulation
was ignited from a point in the northeast corner of the domain $2$ seconds from the start, and the fire perimeter
was recorded after $40$ minutes.  This perimeter  
and ignition location were used to generate an artificial history for the first $40$ minutes,
which was replayed in the second simulation. Therefore, the fire perimeters in both simulations are identical at $40$ minutes. Both simulations were then allowed to advance another $28$ minutes,
using the standard coupled model.  The outputs were then collected for analysis.

\begin{figure*}[ht]
\vspace{-.3in}
\centering
\subfloat [The difference in the winds of the direct simulation minus the artificial propagation at $68$ minutes.]
{
\includegraphics[height=2.5in]{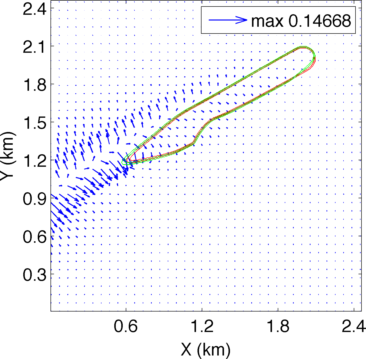}
\label{fig:arrows}
} \hspace{.2in}
\subfloat [The relative error in the speed of the wind at $68$ minutes.] 
{
\includegraphics[height=2.5in]{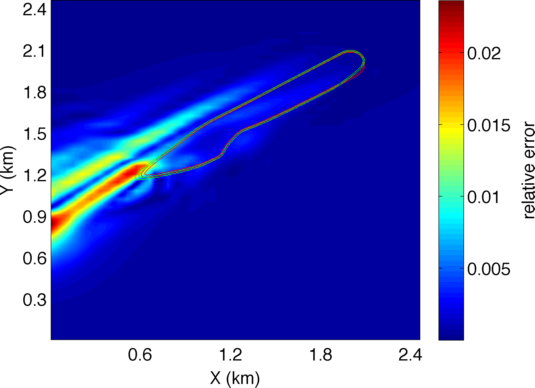}
\label{fig:relerror}
}
\caption{}
\label{fig:diff2d}
\end{figure*}

\begin{figure*}
\vspace{-.3in}
\begin{center}
\subfloat[The artificial fire simulation at $40$ minutes.] {
\includegraphics[height=2.75in]{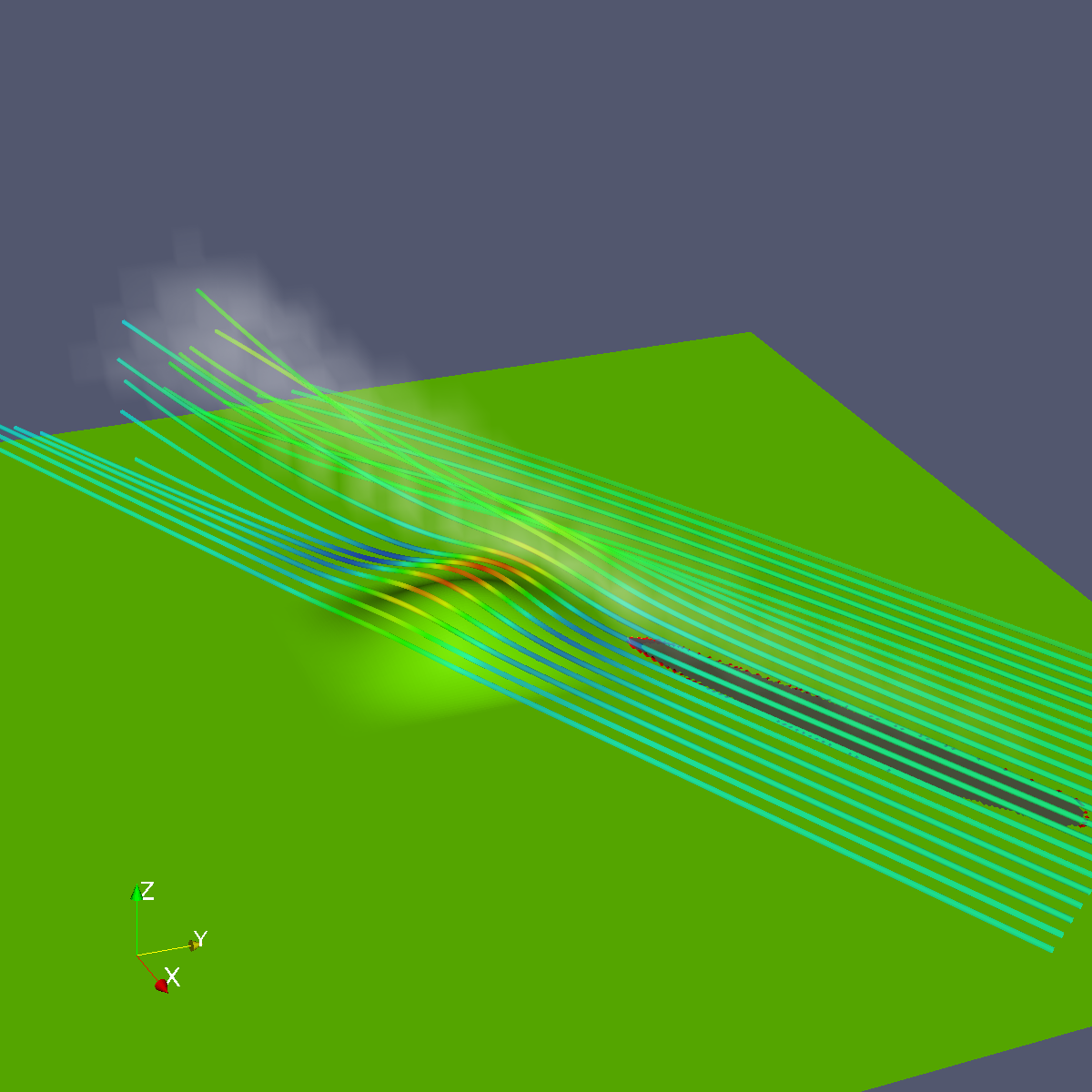}\label{fig:feedback3d}
}\hspace{.2in}
\subfloat[The direct fire simulation at $68$ minutes.] {
\includegraphics[height=2.75in]{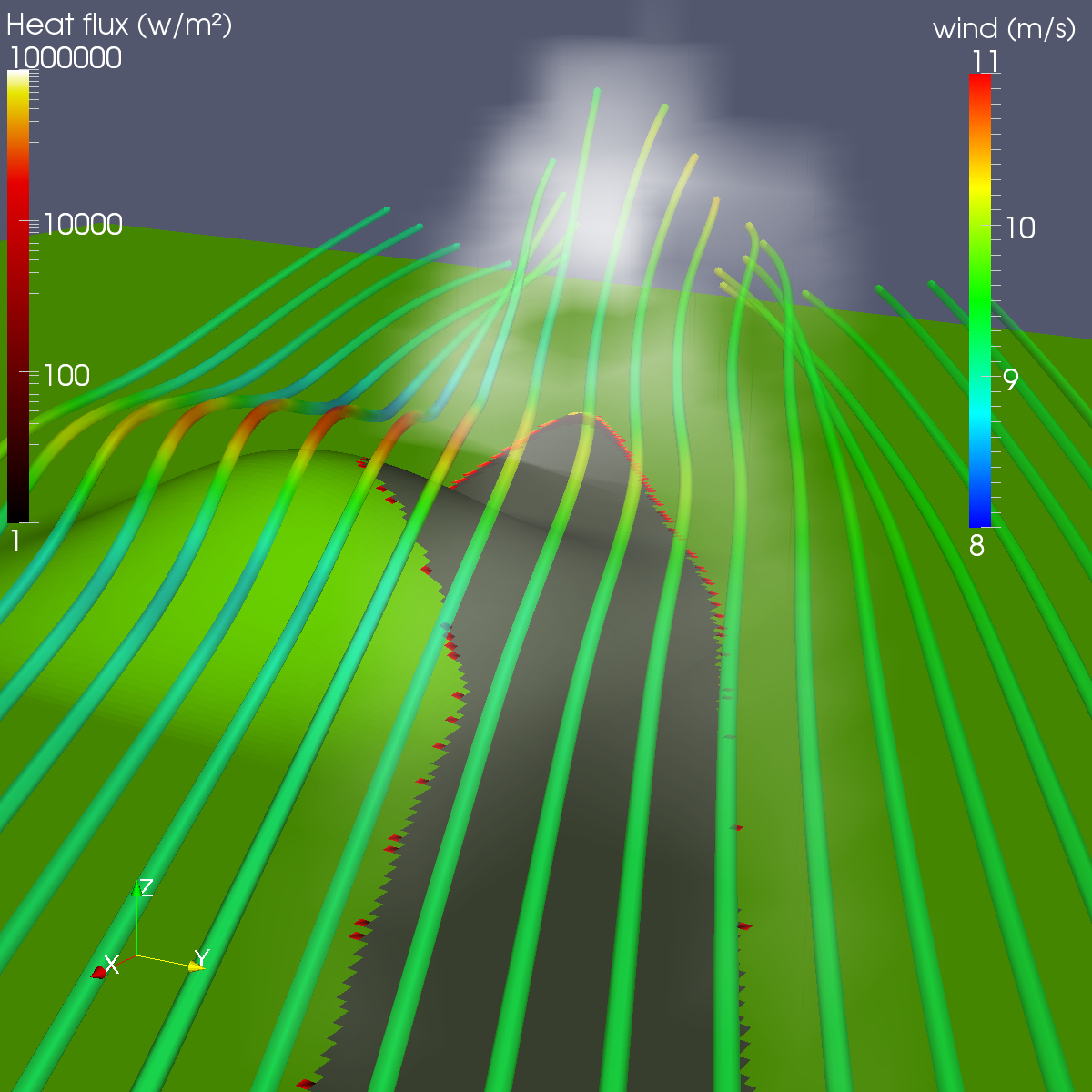}
}
\caption{}\label{fig:vis3d}
\end{center}
\end{figure*}

Any differences in the simulations after this time are a result of the error of the artificial fire propagation.  In Fig.~\ref{fig:vis3d}, we show 
3D renderings of the simulation.  The streamlines near the surface show the updraft created as a result of
the heat output from the fire.  In Fig.~\ref{fig:feedback3d}, the fire is affecting the atmosphere despite being propagated
artificially.  A semi-transparent volume rendering of QVAPOR was added to simulate the smoke release.  In 
Fig.~\ref{fig:diff2d}, the differences in the wind between the two simulations at $68$ minutes and the 
fire perimeter are shown.  Fig.~\ref{fig:arrows}, shows the difference of the 
wind from the direct fire propagation minus the wind from the artificial propagation.  
Fig.~\ref{fig:relerror} shows the relative error
in the wind speed defined as the norm of the difference from Fig.~\ref{fig:arrows}, divided by the wind speed from the direct simulation.
This shows that the maximum error at the end of this $68$ minute simulation is less than $2.5\%$.
In this case, the Froude number is about $F_c=0.79$, showing that the heating from the fire may significantly affect ambient wind, therefore small differences caused by using the artificial history have an effect. The effect is concentrated downwind from the fire, as it could be expected.

\section{Conclusion}

We have presented a parameterization of the fire up to a point in time when a
fire perimeter is given. The parameterization allows for the changes in the
atmospheric circulation to develop, caused by the heat flux from the fire.
This provides appropriate starting conditions for the computation to continue
with the full coupled fire-atmosphere simulation. We have shown on an ideal
example that the differences in the state of the atmosphere between a complete
fire simulation and when the parameterization is used are not significant. In the studied case, the 
coupling between the fire and the atmosphere was strong. It would be interesting to observe how
the the differences change if the problem moves from  the wind-driven ($F_c>1$) to the plume driven regime ($F_c<1$) regime.
This will be studied elsewhere. We plan to study also algorithms for more general domains, not just
star-shaped, and to take into account different rates of fire propagation due to fuel nonhomogeneity. 

\nocite{Mandel-2009-DAW,Mandel-2011-CAF,Clark-2004-DCA,Clark-1996-CAF-a,Sullivan-2009-RWF}
\bibliographystyle{ametsoc}
\bibliography{wrf2011vk,../../references/geo,../../references/other}

\end{document}